\documentclass[aps,twocolumn,prl,superscriptaddress,nofootinbib,showpacs,letterpaper]{revtex4}

\usepackage{amsmath,amssymb}
\usepackage[dvips]{graphicx}
\usepackage{longtable}
\usepackage[usenames]{color}
\usepackage[normalem]{ulem}
\usepackage{bm}

\begin{document}

\newcommand{\comment}[1]{}
\newcommand{\E}{\mathrm{E}}
\newcommand{\Var}{\mathrm{Var}}
\newcommand{\bra}[1]{\langle #1|}
\newcommand{\ket}[1]{|#1\rangle}
\newcommand{\braket}[2]{\langle #1|#2 \rangle}
\newcommand{\mean}[2]{\langle #1 #2 \rangle}
\newcommand{\be}{\begin{equation}}
\newcommand{\ee}{\end{equation}}
\newcommand{\ba}{\begin{eqnarray}}
\newcommand{\ea}{\end{eqnarray}}
\newcommand{\SD}[1]{{\color{magenta}#1}}
\newcommand{\rem}[1]{{\sout{#1}}}
\newcommand{\alert}[1]{\textbf{\color{red} \uwave{#1}}}
\newcommand{\Y}[1]{\textcolor{blue}{#1}}
\newcommand{\R}[1]{\textcolor{red}{#1}}
\newcommand{\B}[1]{\textcolor{black}{#1}}
\newcommand{\C}[1]{\textcolor{cyan}{#1}}
\newcommand{\db}{\color{darkblue}}
\newcommand{\aaron}[1]{\textcolor{cyan}{#1}}
\newcommand{\ac}[1]{\textcolor{cyan}{\sout{#1}}}
\newcommand{\intinfty}{\int_{-\infty}^{\infty}\!}
\newcommand{\Tr}{\mathop{\rm Tr}\nolimits}
\newcommand{\const}{\mathop{\rm const}\nolimits}

\title{Branching of quasinormal modes for nearly extremal Kerr black holes}

\author{Huan Yang}
\affiliation{Theoretical Astrophysics 350-17, California Institute
of Technology, Pasadena, CA 91125, USA}
\author{Fan Zhang}
\affiliation{Theoretical Astrophysics 350-17, California Institute of Technology, Pasadena, CA 91125, USA}
\author{Aaron Zimmerman}
\affiliation{Theoretical Astrophysics 350-17, California Institute of Technology, Pasadena, CA 91125, USA}
\author{David A.\ Nichols}
\affiliation{Center for Radiophysics and Space Research, Cornell University, 
Ithaca, New York 14853, USA}
\author{Emanuele Berti}
\affiliation{Department of Physics and Astronomy, The University of Mississippi, University, MS 38677, USA}
\affiliation{Theoretical Astrophysics 350-17, California Institute of Technology, Pasadena, CA 91125, USA}
\author{Yanbei Chen}
\affiliation{Theoretical Astrophysics 350-17, California Institute
of Technology, Pasadena, CA 91125, USA}

\begin{abstract}
We show that nearly extremal Kerr black holes have two distinct sets
of quasinormal modes, which we call zero-damping modes (ZDMs) and
damped modes (DMs).  The ZDMs exist for all harmonic indices $l$ and
$m \ge 0$, and their frequencies cluster onto the real axis in the
extremal limit.  The DMs have nonzero damping for all black hole
spins; they exist for all counterrotating modes ($m<0$) and for
corotating modes with $0\leq \mu\lesssim \mu_c=0.74$ (in the eikonal
limit), where $\mu\equiv m/(l+1/2)$. When the two families coexist,
ZDMs and DMs merge to form a single set of quasinormal modes as the
black hole spin decreases.  Using the effective potential for
perturbations of the Kerr spacetime, we give intuitive explanations
for the absence of DMs in certain areas of the spectrum and for the
branching of the spectrum into ZDMs and DMs at large spins.
\end{abstract}


\maketitle

\begin{figure}
\includegraphics[width=0.48\textwidth]{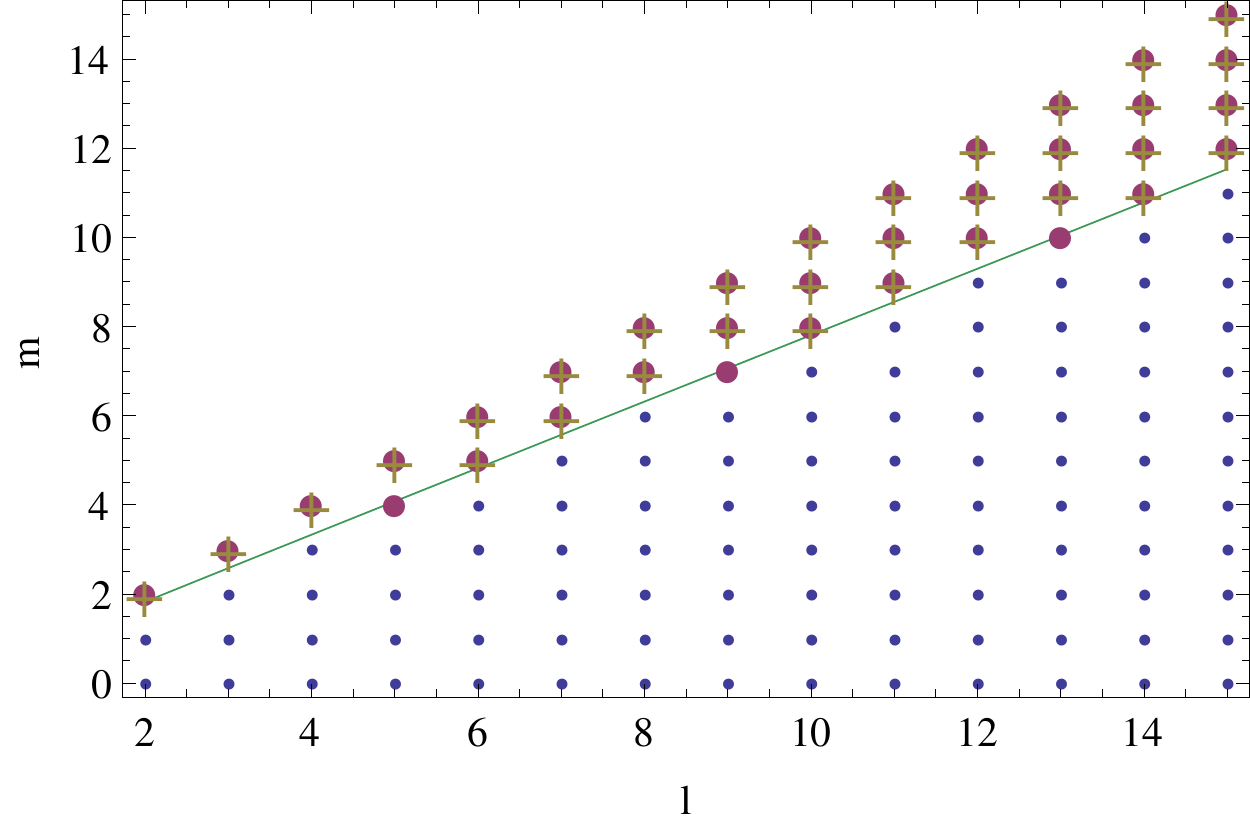}
\caption{(Color online.) Phase diagram for the separation between the single- and double-branch regime for NEK BHs. Large purple dots and gold crosses correspond to $(l,m)$ pairs with only ZDMs for perturbations with spin $-2$ and $0$, respectively. Smaller blue dots correspond to $(l,m)$ pairs with both ZDMs and DMs. The green line is the phase boundary, computed using the eikonal approximation.
\label{fig:phase}}
\end{figure}

\noindent{\bf{\em I.~Introduction.}}~Nearly extremal Kerr (NEK) black holes 
(BHs)---i.e., BHs for which the dimensionless angular momentum $a \approx 1$ in 
the geometrical units, $G = c = M =1$, used in this paper---have drawn much
attention recently. 
Besides the mounting evidence for fast-rotating BHs in astronomy
\cite{astro_evidence}, NEK BHs have considerable theoretical significance, 
e.g., in studies of weak cosmic censorship \cite{censor} and in calculations 
of black-hole entropy \cite{strominger}. 

For extremal Kerr BHs ($a=1$) the near-horizon geometry reduces to 
AdS$_2\times$ S$^2$ \cite{bardeen}. 
This observation led to the Kerr/CFT conjecture, which states that extremal 
Kerr BHs are dual to the chiral limit of a two-dimensional conformal field 
theory \cite{kerrcft}. In the past few years the extremal Kerr spacetime and spacetimes violating the Kerr bound were shown to be unstable \cite{instability}. The stability of BHs depends on the sign of the imaginary part of their complex free vibration modes, called quasinormal modes (QNMs) \cite{Berti:2009kk}. Therefore the NEK QNM frequencies studied here can shed light on the onset of extremal Kerr instabilities and prove useful in quantum field theory (for example, in the calculation of two-point functions \cite{Bredberg:2009pv}). 

Detweiler first used an approximation to the radial Teukolsky
equation for NEK BHs (see also \cite{teukolsky}) to show that QNMs
with angular indices $l=m$ have a long decay time \cite{detweiler}.
Using Detweiler's result, Sasaki and Nakamura \cite{sasaki} calculated
QNM frequencies analytically and Andersson and Glampedakis proposed long-lived
emission from NEK BHs \cite{andersson}.  However, there remains a
long-standing controversy in the literature about what set of QNMs
decay slowly \cite{hod2}, whether long-lived radiation is possible
\cite{cardoso}, and whether the imaginary part of the QNM frequencies
vanishes as $a \to 1$ (compare \cite{sasaki,cardoso} with \cite{hod2}). 
Despite the importance of this problem, our present understanding of the QNM spectrum of NEK BHs is inconclusive.

In a recent paper \cite{yang}, some of us used a WKB analysis to relate Kerr 
QNMs in the eikonal limit to spherical photon orbits around Kerr BHs. 
We pointed out that a subset of spherical photon orbits of extremal Kerr BHs reside
on the horizon and that the corresponding QNMs have zero damping. 
This happens when the parameter $\mu\equiv m/(l+1/2)\gtrsim \mu_c\simeq 0.74$. Hod \cite{hod3} computed $\mu_c$ in the eikonal limit, finding an approximate analytical result in agreement with \cite{yang}.

In this work, we will show that the NEK geometry has two distinct sets of QNMs:
zero-damping modes (ZDMs) and damped modes (DMs). 
ZDMs are associated with the near-horizon geometry of the BH, and they exist 
for {\it all} allowed values of $l$ and $m\ge 0$ (we classify modes using 
Leaver's conventions \cite{leaver}, but we use units in which the 
BH has mass $M=1$). 
DMs are associated with peaks of the potential barrier; in the eikonal
limit, they exist when $\mu \leq 0.74$.  This implies that ZDMs and
DMs coexist if $0\leq \mu \leq 0.74$. Figure~\ref{fig:phase} is a
``phase diagram'' in QNM space, showing the regions where either the ZDMs or both the DMs and the ZDMs exist for scalar and gravitational perturbations with
$l \leq 15$. We will discuss this phase diagram further below.
When the ZDMs and DMs coexist, and when the BH spin $a$ is small, for each $(l,\,m)$ there is only a single set of QNMs characterized by the overtone number $n$ (where modes with larger $n$ have stronger damping). 
For larger $a$, this set of QNMs appears to break into two branches. The DM branch originates from lower-overtone modes at smaller $a$, and its modes retain a finite decay rate as $a \to 1$. The ZDM branch originates from higher-overtone modes whose imaginary part becomes smaller than that of DMs as $a \to 1$, thereby forming the second branch. This is similar to the behavior of eigenmodes in quantum mechanics when we parametrically split a single potential well into two potential wells (cf.~Fig.~\ref{fig:pplot} below, as well as \cite{ChandraFerrari} for a somewhat analogous phenomenon in the theory of oscillations of ultracompact stars).

\vspace{0.2cm}

\noindent{\bf{\em II.~Matched expansions.}}
For $\epsilon\equiv 1-a \ll 1$
and $\omega-m/2 \ll 1$, the radial Teukolsky equation can be written in a 
self-similar form when $(r-1) \ll 1$ and in an asymptotic form (by setting $a=1$)
when $(r-1) \gg \sqrt{\epsilon}$ (cf. \cite{teukolsky,detweiler,YangInPrep}). 
The solutions of the Teukolsky equation in these regions
(hypergeometric and confluent hypergeometric functions, respectively)
can be matched at $\sqrt{\epsilon} \ll (r-1) \ll 1$ to provide the
following condition for QNM frequencies:
\begin{align}
\label{eqdetweiler}
e^{-\pi \delta-2 i \delta \ln(m)-i\delta \ln(8\epsilon)}\frac{\Gamma^2(2 i \delta)\Gamma(1/2+s-i m -i \delta)}{\Gamma^2(-2 i \delta)\Gamma(1/2+s -i m +i \delta)}
\notag \\
\times\frac{\Gamma(1/2-s-i m -i \delta)\Gamma[1/2+i(m-\delta-\sqrt{2}\tilde{\omega})]}{\Gamma(1/2-s-i m +i \delta)\Gamma[1/2+i(m+\delta-\sqrt{2}\tilde{\omega})]}
=1 .
\end{align}
Here we denote the eigenvalues of the angular Teukolsky equation by 
${}_sA_{lm}$, and we define $\delta^2 \equiv 7m^2/4-(s+1/2)^2-{}_sA_{lm}$ and
$\tilde{\omega} \equiv (\omega-m\Omega_H)/\sqrt{\epsilon}$ [note that
$\Omega_H=a/(r^2_+ +a^2)$ is the horizon frequency and 
$r_{+}=1+ \sqrt{1-a^2}$ is the horizon radius].
Scalar, electromagnetic, and gravitational perturbations correspond to spin 
$s=0,-1,-2$, respectively. 
If we choose the conventions that $\mathcal{R}e(\delta) \ge 0$ and
$\mathcal{I}m(\delta) \ge 0$ when $\delta^2$ is positive and negative, 
respectively, then the left-hand side of Eq.~\eqref{eqdetweiler} is usually a 
very small number, except when it is near the poles of the $\Gamma$-functions 
in the numerator. 
When $m \ge 0$, we can always find the solution near the poles at negative 
integers:
\be\label{eqhod}
1/2+i(m-\delta-\sqrt{2}\tilde{\omega}) \approx -n ,
\ee
or
\be\label{eqhod2}
\omega \approx \frac{m}{2}-\frac{\delta \sqrt{\epsilon}}{\sqrt{2}}-i\left (n+\frac{1}{2}\right ) \frac{\sqrt{\epsilon}}{\sqrt{2}} .
\ee
Note that the overtone index $n$ of these ZDM frequencies need not correspond 
precisely to the same overtone index of Kerr QNMs at lower spins.
This set of solutions was first discovered by Hod \cite{hod2}. 
The matched-expansion derivation shows that this set of modes
depends on the near-horizon region of the Kerr BH. 
Equation \eqref{eqhod2} is quite accurate when $|\delta|\gg 1$, but when 
$|\delta|<1$ it needs an additional correction \cite{YangInPrep}.
However, the $\sqrt{\epsilon}$ scaling of the decay rate is still correct when
$|\delta|<1$. 
The solutions to Eq.~\eqref{eqhod2} with $m < 0$ are those that arise from the
symmetry $\omega_{l,m}=-\omega^*_{l,-m}$; there are no solutions with $m <0$ 
and $\mathcal{R}e(\omega)>0$, when $\omega -m/2$ is not small.
Thus, the ZDMs only exist in the {\it corotating} regime $m\geq 0$.

Another set of solutions of Eq.~\eqref{eqdetweiler} may exist
when $\delta^2<0$ and $2 i \delta \approx -n$, with $n$ a positive
integer.
A more detailed analysis shows that, in this case, two nearly
degenerate hypergeometric functions have comparable contribution to
the near-horizon solution \cite{YangInPrep}.  As a result,
Eq.~(\ref{eqdetweiler}) is no longer valid when $2 i \delta \approx
-n$.  As a consistency check, we looked for solutions with $2 i \delta
\approx -n$ using Leaver's method and we did not find any.

\begin{figure}[t!]
\includegraphics[width=1.0\columnwidth]{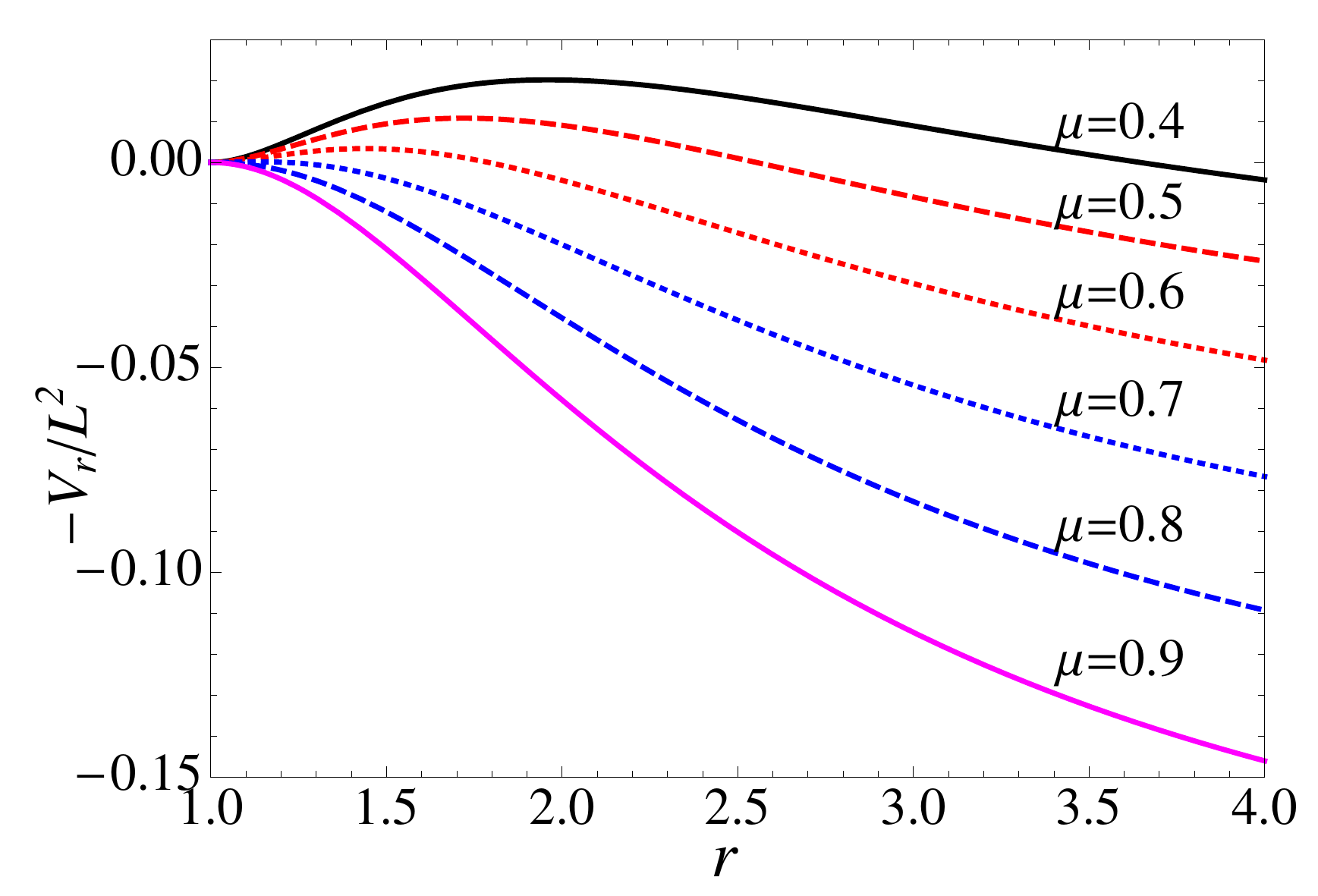}
\caption{(Color online.) Plot of the potential term for $a=1$, Eq.~(\ref{potential}), for $\mu=0.4$, $0.5$, $ 0.6$, $0.7$, $0.8$, and $0.9$ (black-solid, 
red-dashed, red-dotted, blue-dotted, blue-dashed, and magenta-solid curves, 
respectively).
The transition from single-branch to double-branch happens between $\mu=0.7$ and $\mu=0.8$. }
\label{fig:pplot}
\end{figure}

\vspace{0.2cm}

\noindent{\bf{\em III.~WKB analysis.}}~The matched-expansion method
assumes that $\omega \approx m/2$, but Eq.~\eqref{eqdetweiler} does
not hold for modes which do {\it not} meet this requirement
(i.e., DMs).  To compute these modes, we will instead use a WKB
analysis in the eikonal limit $l\gg 1$.
The radial Teukolsky equation when $l\gg 1$ is \cite{yang}
\begin{subequations}
\begin{equation}
\label{eqr}
\frac{d^2 u_r}{d r^2_*}+V_r u_r=\frac{d^2 u_r}{d r^2_*}+\frac{K^2-\Delta\lambda^0_{l m}}{(r^2+a^2)^2} u_r=0 \,,
\end{equation}
with
\begin{align}
K&=-\omega(r^2+a^2)+am, \quad \frac{d}{dr_*} \equiv \frac{\Delta}{r^2+a^2}\frac{d}{dr}, \nonumber \\
\lambda^0_{l m} &=A_{lm}+a^2\omega^2-2am\omega, \quad \Delta = r^2 - 2  r +a^2 \,.
\label{eqexplan}
\end{align}
\end{subequations}
We define $\omega \equiv \omega_R-i \omega_I$, and we note that the real and
imaginary parts scale as $\omega_R \propto l$ and $\omega_I \propto l^0$, 
while the angular constant scales as $A_{lm} \propto l^2$. 
We only keep the leading-order terms in the eikonal limit in the following
discussion (therefore all $s$-dependent terms are neglected, and the $A_{lm}$ 
are real). In Fig.~\ref{fig:pplot} and below, we will refer to $-V_r$ as ``the potential''.
According to the WKB analysis and its geometric correspondence in \cite{yang}, 
the position of the peak of the potential asymptotes the horizon as 
$a \rightarrow 1$ for some of the corotating modes. 
For this set of QNMs, one can verify that $V''_r$ (where primes denote derivatives 
with respect to $r_*$) scales as $\Delta^2$; thus, the peak $r_0$ of the 
potential becomes broad as $r_0$ approaches the horizon. 
It then follows that $\omega_I \propto \sqrt{V''_r}/\partial_\omega V_r \to 0$,
and $\omega_R \rightarrow m/2$ in order to satisfy $V_r(\omega_R, r_0)=0$ for 
these modes.
Assuming that $r_0 = 1 + c \sqrt{\epsilon}$ for the nearly extremal
modes, where $c$ is some constant, we can apply the eikonal equations
in \cite{yang} and obtain
\begin{align}
\label{eqscaling}
r_0 \approx 1+\frac{ m\sqrt{2 \epsilon}}{\mathcal{F}_0},\ \omega_R \approx \frac{m}{2}-\frac{\mathcal{F}_0 \sqrt{\epsilon} }{\sqrt{2}},\ \omega_I \approx \left(n+\frac{1}{2}\right)\frac{\sqrt{\epsilon}}{\sqrt{2}} \,,
\end{align}
with $\mathcal{F}_0 = \sqrt{7m^2/4 - A_{lm}(\omega=m/2)}$. 
Comparing this result with Eqs.~(\ref{eqhod2}) and \eqref{eqscaling}, we can see the 
two sets of frequencies are essentially the {\it same} modes, although obtained in very different ways. 
Here $\mathcal{F}_0^2$ and $\delta^2$ differ by $1/4$, which is reasonable 
because in the eikonal limit $\mathcal{F}_0 \propto l$ and $\delta \propto l$ 
(making $1/4$ a higher-order correction). 

To build intuition about $\mathcal{F}_0$ and $\delta$, we look at $V_r$ for extreme Kerr BHs, with $\omega$ replaced by $m/2$: 
\begin{equation}
\label{potential}
V_r=L^2\frac{(r-1)^2}{(r^2+1)^2}\left [ \frac{(r+1)^2}{4}\mu^2-\alpha(\mu)+\frac{3}{4}\mu^2\right ] ,
\end{equation}
where $L \equiv l+1/2$ and $\alpha(\mu) \equiv A_{lm}/L^2$. 
According to the WKB analysis of the radial Teukolsky equation \cite{Iyer}, 
the QNM frequencies are determined by the peak of the potential.
As shown in Fig.\ \ref{fig:pplot}, when $\mu$ is large the 
maximum of the potential is at the horizon, $r=1$, as expected for ZDMs. 
As $\mu$ decreases and falls below some critical value $\mu_c$, the 
peak moves outside the horizon, and the horizon becomes a local minimum of the
potential. At the peak $\omega_I$ is nonzero because  
$d^2 V_r/d r^2_*|_{r_0} \neq 0$, so we have DMs.
The criterion for having no peak outside the horizon is
\begin{equation}\label{eqalpha}
 \frac{(r+1)^2}{4}\mu^2-\alpha(\mu)+\frac{3}{4}\mu^2 >0\quad {\rm for} \quad r=1\,,
 \end{equation}
i.e., $\mathcal{F}_0^2 >0$ (or $\delta^2>0)$. 
The values at which $\mathcal{F}_0^2$ (or $\delta^2$) vanish lead to the 
condition for the critical $\mu_c$: $\alpha(\mu_c)=\frac{7}{4}\mu_c^2$.
If we use the approximation 
$\alpha(\mu)\approx 1-a^2\omega^2(1-\mu^2)/(2L^2)$ \cite{yang}, 
this will reproduce Hod's approximate analytical result 
$\mu_c \approx [(15-\sqrt{193})/2]^{1/2}$ \cite{hod3}. 
We can obtain the exact $\mu_c$ (in the eikonal limit) by inserting 
$\alpha(\mu_c)=\frac{7}{4}\mu_c^2$ into the Bohr-Sommerfeld condition for 
$\alpha$ derived in \cite{yang}:
\begin{equation}
\int^{\theta_+}_{\theta_-}\sqrt{\alpha-\frac{\mu^2}{\sin^2\theta}+\frac{\mu^2}{4}\cos^2\theta}=(1-|\mu|)\pi,
\end{equation}
where $\theta_+=\pi-\theta_-$ and $\theta_-=\arcsin(\sqrt{3}-1)$ are the angles
at which the integrand vanishes. 
Therefore we have
\begin{equation}\label{eqeikoncon}
\mu_c =\frac{1}{1+\mathcal{I}/\pi}, \quad \mathcal{I}=\int^{\theta_+}_{\theta_-}d\theta\sqrt{\frac{7}{4}-\frac{1}{\sin^2\theta}+\frac{1}{4}\cos^2\theta},
\end{equation}
which yields the numerical value $\mu_c\simeq 0.74398$.  In the
eikonal limit, when $\mu>\mu_c$ NEK BHs have only ZDMs
(``single-phase regime'') ; when $0\le\mu\le\mu_c$, both DMs and ZDMs
exist (``double-phase regime'').

\vspace{0.2cm}

\noindent{\bf{\em IV.~Phase boundary.}} Although there is a clear criterion for
determining the boundary between the single-phase regime and the double-phase 
regime in the eikonal limit (when $\mu<\mu_c$, the peak of the potential no
longer resides on the horizon) it is not immediately clear if a similar 
criterion holds when $l$ is small.
For scalar perturbations, however, we can write
the radial Teukolsky potential for extreme-Kerr BHs with generic $l,m$, under 
the assumption that $\omega = m/2$ (and, therefore, the ${}_0A_{lm}$ remain 
real for the ZDMs):

\begin{align}
V_r = &\frac{(r-1)^2}{(r^2+1)^2}\left [ \frac{(r+1)^2}{4}m^2-{}_0A_{lm}  \right ]\nonumber \\
&+\frac{(r-1)^2}{(r^2+1)^2}\left [\frac{3}{4}m^2+\frac{(r-1)(2r^2+3r-1)}{(1+r^2)^2} \right ] .
\end{align}

It is not difficult to see that there is still no peak outside
the horizon when $\frac{7}{4}m^2>{}_0A_{lm}$, or $\mathcal{F}_0^2>0$. 
For electromagnetic and gravitational perturbations the potential terms $-V_r$ 
are complex functions, thereby making the positions of their extrema more 
difficult to define. 
Detweiler \cite{detweiler2}, however, has shown that the radial function can be
transformed so that it satisfies a differential equation with a real potential.
Using this potential, the criterion to exclude peaks outside the horizon is
\cite{YangInPrep}: 
\be
\label{eqcriall}
\mathcal{F}^2_s\equiv \frac{7}{4}m^2-s(s+1)-{}_s A_{lm}\left (\omega=\frac{m}{2}\right )>0 .
\ee
Note that this expression respects the pairing symmetry 
${}_{-s}A_{lm}={}_sA_{lm}+2s$, and that for all $s$, $\mathcal{F}^2_s$ and 
$\delta^2$ differ from each other only by $1/4$. 
For $s=0,-2$ and $2 \le l \le 100$, we have searched all QNMs numerically and 
have not found any mode simultaneously satisfying $\delta^2<0$ and 
$\mathcal{F}_s^2>0$; therefore, the sign of $\delta^2$ also determines whether 
a peak exists outside the horizon. 
In addition, we have used Leaver's continued-fraction algorithm to determine the 
phase boundary numerically. 
As shown in Fig.~\ref{fig:phase}, the actual phase boundary matches the 
criterion predicted by the eikonal limit, $\mu=\mu_c$. 
In addition, for scalar and gravitational perturbations, we find numerically that modes are in the single-phase regime when $\mathcal{F}_s^2>0$ for all $l\leq 15$. 
This reinforces our physical understanding that DMs are associated with a peak
of the potential outside the horizon, while ZDMs are somewhat similar in nature to the $s$-modes in ultracompact stars \cite{ChandraFerrari}.

\vspace{0.2cm}

\noindent{\bf{\em V.~Bifurcation.}}~Schwarzschild and slowly spinning Kerr BHs 
have a single set of QNMs for each $l,m$ that are characterized by their 
overtone number $n$.
If the ZDMs originate from modes at higher-$n$ than the DMs when the BH spin is low, then when the spin increases beyond a critical value $a_s=1-\epsilon_s$, a single set of QNMs may split into two branches.
 
\begin{figure}[t!]
\includegraphics[width=1.0\columnwidth]{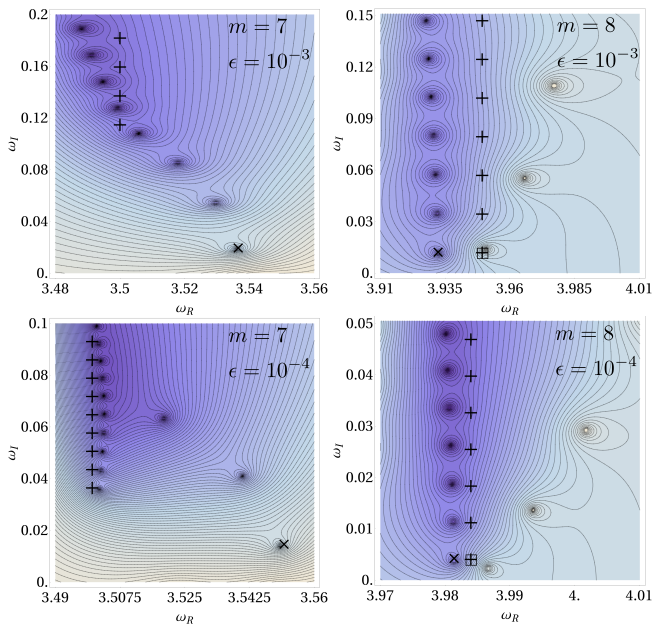}
\caption{(Color online.) QNM frequencies with $l=10$ for NEK BHs. 
Contours are constant values of the logarithm of the continued fraction in the 
complex plane; darker shading indicates values near zero. 
The $+$ symbols are the ZDM predictions, a $\times$ is the lowest-overtone WKB 
prediction from \cite{yang}, and the box is centered at the WKB prediction from
Eq.~\eqref{eqscaling}. 
No branching is observed for modes with $m=8$. 
Note that the closed contours with light shading have large values and do not 
correspond to any QNM. Further discussion of the figure is in the text.}
\label{fig:cplot}
\end{figure}

We numerically investigate this bifurcation effect by examining the complex 
QNM frequency plane to search for solutions of Leaver's continued-fraction 
equations \cite{leaver,YangInPrep}.
In Fig.~\ref{fig:cplot}, we plot the contours of constant value of the 
logarithm of the continued-fraction expansion, truncating at $N=800$ terms. 
The QNM frequencies correspond to the local minima of this sum, where the contours cluster. The shading indicates the value of the fraction, with darker values nearly zero.

When $\mu < \mu_c$, a single set of QNMs splits into two branches for increasing $a$ (see the left-hand panels of Fig.~\ref{fig:cplot}, where $l=10,\, m=7$, as the spin increases from $a= 0.9990$ to $a=0.9999$ from the upper panel to the lower). The ZDM branch is quite accurately described by Eq.~(\ref{eqhod2}); the imaginary part of the ZDMs scales like $\sqrt{\epsilon}$, and they move towards the real axis as $\epsilon\to 0$. The DM branch changes relatively little with increasing spin (it is expected that the WKB peak can only support a finite number of modes \cite{YangInPrep}, and there are only 3 DMs in the lower-left panel). In this case, the WKB formulae of~\cite{yang} are in good agreement with the lowest-overtone DM
(marked with a $\times$ in the figure).
 
For $\mu > \mu_c$ there is no bifurcation, and the modes are predicted fairly 
well by Eq.~\eqref{eqhod2}. 
We can see this in the right-hand panels of Fig.~\ref{fig:cplot}, where 
$l=10, \, m=8$ and we again raise the spin from $a=0.9990$ to $a=0.9999$. 
For the $m=8$ modes, we also mark the leading-order WKB prediction of 
Eq.~\eqref{eqscaling} with a box.
For the bifurcation effect, we can define a benchmark $a_c=1-\epsilon_c$ as
the BH spin at which the imaginary part of the {\it fundamental} ZDM is equal to that of the {\it fundamental} DM:
\be
\label{eqcrita}
\frac{\sqrt{\epsilon_c}}{2\sqrt{2}}(1+2|\delta|)=\left. \frac{1}{2}\frac{\sqrt{2V''_r}}{\partial_{\omega}V_r}\right |_{r_0}.
\ee
The right-hand side of Eq.~(\ref{eqcrita}) can be evaluated using the 
approximate WKB formula in \cite{yang}. 
Since both sides of Eq.~\eqref{eqcrita} depend on $\epsilon$, we solve for 
$\epsilon_c$ iteratively; this converges quickly for a variety of initial spins.
By computing $\epsilon_c$ for $l \leq 15$ and $0<m<(l+1/2)\mu_c$, we find that 
$L^2 \epsilon_c = 10^{-3}(11.6 - 3.12 \mu - 18.0 \mu^2)$ is a reasonable 
fitting formula. 
For the $l=10,\,m=7$ case, Eq.~(\ref{eqcrita}) gives $\epsilon_c \sim10^{-5}$, 
which is in agreement with numerical results; for the $l =2,\, m=1$ case it 
gives $\epsilon_c \sim 10^{-3}$. 

In Fig.~\ref{fig:cplot}, however, it is clear that the bifurcation actually 
starts when the {\it fundamental} ZDM's imaginary part equals the imaginary part 
of the {\it highest-overtone} DM (in Fig.~\ref{fig:cplot} it is the third overtone).
This happens at a spin $a_s<a_c$. 
Because we do not have a good estimate of the number of modes in the DM branch
(beyond the fact that it should be proportional to $L$ and a function of $\mu$ in the eikonal limit \cite{YangInPrep})
and because WKB techniques are not accurate for these high-overtone DMs, finding an analytic solution for $a_s$ remains an open problem.

\vspace{0.2cm}

\noindent{\bf{\em VI.~Conclusions.}}~We identified two different regimes in the
NEK QNM spectrum. 
In the double-phase regime, we found that the lowest ZDM becomes less damped 
than the lowest DM at some critical $a_c$, for which we provided an analytical estimate. 
For sufficiently large $a$, Eq.~(\ref{eqhod2}) is accurate at the least for those ZDMs with smaller decays than the point where the branches bifurcate. We estimate that the number of ZDMs below the bifurcation is $\propto \sqrt{\epsilon_s/\epsilon}$ \cite{YangInPrep}. In the future, we would like to investigate the behavior of the ZDM branch in the high-overtone limit \cite{uri}, where these approximations break down. 

\vspace{0.2cm}

\noindent{\bf{\em VII.~Acknowledgments.}}~We thank Sam Dolan for advice on the WKB method, and Zhongyang Zhang for discussions during the early stages of this work. This research is funded by NSF Grants PHY-1068881 and PHY-1005655, CAREER Grants PHY-0956189 and PHY-1055103, NASA Grant No.NNX09AF97G, the Sherman Fairchild Foundation, the Brinson Foundation, and the David and Barbara Groce Startup Fund at Caltech.

\end{document}